\documentstyle[eqsecnum,aps]{revtex}
\newcommand{\lsim}
{\raisebox{-0.7ex}{$\;\stackrel{\textstyle <}{\textstyle\sim}\;$}}

\begin{document}

\title{
Equilibrial Charge of Grains and Low-temperature Conductivity of 
Granular Metals.  }

\author{
E.M.
Baskin, M.V. Entin }
\address{
Institute of Semiconductor Physics, Russian Academy of Sciences,
Siberian Branch,
RUSSIA,\\ E-mail:Entin@isp.nsc.ru}
\maketitle

\begin{abstract}
The low-temperature equilibrial state of a system of small metal
grains, embedded into insulator, is studied. We find, 
that the grains may be charged due to the fluctuations of the surface 
energy of electron gas in grains, rather than quantization of 
electron states.  The higherst-occupied level in a grain fluctuates 
within the range of order of charging energy below the overall 
chemical potential. The system, called a gapless Hubbard insulator,  
has no overall energy gap, while the transfer of an electron on 
finite distances costs finite  energy. The ionization energy 
is determined mostly by the intragrain Coulomb repulsion, rather than 
a weak intergrain interaction, responsible for the Coulomb gap. 

The hopping  transport in the system is studied. The hopping energy
is determined by the charging energy. At low temperature the
transport has gapless character.
 \end{abstract}


\narrowtext
\twocolumn
The hopping conductivity of metal-insulator 
composite was intensively studied in the past decades\cite{Abel,3,4}. 
The numerous experiments evidenced  typical behavior of conductivity 
$\sigma\sim \exp -(T_0/T)^{1/2}$, which needs zero energy gap of the 
system \cite{1,Pollak}. It is generally accepted that the energy 
scale of hopping transport originates from the Coulomb interaction.  
Two main reasons were put forward, intra- and intergrain e-e 
repulsion.  The first one leads to  a Hubbard-like hard gap and 
constant activation energy of hopping conductivity, the second is 
responsible for the smooth Coulomb gap.  The Coulomb gap requires 
grains charging at low temperature. At the same time the origin of 
charging is not clear.  In \cite{1}, \cite{Pollak} the electrons 
redistribution was ascribed to the difference of their work 
functions.  This assumption is reasonable relative to the grains of 
different metals, but this is not the case in the most experiments.  
Chui \cite{Chui} assumed that the fluctuations of level spacing may 
overweight the charging energy, leading to grains ionization.

Another reason of charging was proposed in\cite{Ortuno}.  The
numerical study of \cite{Ortuno} showed that the large portion of
grains should be charged.  The charging
was attributed to the large variation of the highest-occupied
energy level due to the wide fluctuation of level spacing both in
integrable and non-integrable systems.

The system of doped  semiconductor quantum dots was studied in our
work \cite{5}.  In this system the dots becomes charged due to
fluctuations of local chemical potential, caused by the fluctuations
of impurity numbers. This system was evidenced to be a gapless
Hubbard insulator.

The purpose of the Letter is the study of low-temperature equilibrial 
states and hopping transport in a system of large metal grains, 
embedded in insulator. Unlike to other studies, we want account 
for the electron quantization in the first order, responsible 
for the grain surface energy.  We argue, that  grain charging  at 
low-temperature results from variations of surface energy of electron 
gas in different grains.  We evidenced, that in dependence on the 
ratio of Coulomb and Fermi energy, the mean charge may exceed the 
electron charge, producing zero-gap Hubbard insulator.  Then we shall 
study the hopping transport in  this system.

Let us discuss the typical parameters of the problem. The charging of 
a grain with capacitance $C$ by an electron costs the energy
$U_c=e^2/2C$. This energy is less than the energy of e-e interaction 
on the mean distance between electrons inside grain $U_0$, which in 
most cases is less, than the Fermi energy of bulk metal $E_F$.  Hence 
inside grain the electron gas is ideal, while  the charging 
energy $U_C$ determines  transport properties. 

The other important parameter is  interlevel distance 
$\delta E\sim (\nu V)^{-1}$, where $\nu\propto E_F^{1/2}$ is the 
density of states of bulk metal, $V$ is a grain volume. Since 
$C\propto V^{1/3}$, $U_c\gg \delta E$ in a grain, larger than the 
atomic size.  At the same time, the surface energy per electron, 
found below, is proportional to $E_FS/V\propto V^{-1/3}$  and remains 
comparable with $U_C$ if $V\to \infty$. So we shall account for 
surface energy and neglect the interlevel distance.  We shall neglect 
the collectivization of electron states in different grains also.

Let us consider a metal-insulator composite containing the metal
grains with small density $N$, $NV\ll 1$, embedded into insulator 
with the permittivity $\kappa$. Within our model, the grains can 
exchange by electrons. 

The existence of surfaces leads to
the surface energy
$\alpha S$, proportional to the surface $S$,  additional with respect
to the infinite volume.
To find the surface tension $\alpha$, consider the $\Omega$-potential
of free electron gas in a neutral metal grain. Since $\alpha$ does not
depend on the shape of grain, one can consider
a rectangular box. In the limit of  large grain size and zero
temperature
\begin{eqnarray} \label{200}
\Omega=-\frac{m^{3/2}(2\mu)^{5/2}}{15\pi^2\hbar^3}V+
\frac{m\mu^2}{8\pi\hbar^2}S,
\end{eqnarray}
where $m$
is electron mass, $\mu$ is the local chemical potential.
According to (\theequation), the coefficient of surface tension
$\alpha$ of ideal electron gas is
$\alpha=\frac{m\mu^2}{8\pi\hbar^2}$.
The
ratio of surface energy to the volume energy is determined by the
ratio of Fermi wavelength to the system size.

The single-electron spectrum strongly depends on the integrability of 
the system determined by the grain shape.  The  $\Omega$-potential 
(\ref{200}) does not depend on the shape of grain with fixed volume 
$V$ and surface $S$; hence $\Omega$ is not sensitive to the 
energy spectrum. 

For neutrality of a  grain the number of electrons  $ 
Z_i=-\frac{\partial \Omega_i}{\partial 
\mu_i}=\frac{(2m\mu_i)^{3/2}}{3\pi^2\hbar^3}V_i-\frac{m\mu_i}{4\pi\hbar^2}S_i$
 should be equal to the number of ions $z_i=nV_i$, where $n$ is the 
density of ions, subscript $i$  numbers  grains. To be 
so, the chemical potential should depend on the number of a grain by 
means of ratio $S_i/V_i$.  The equalibration of electrochemical 
potentials $\mu_i-e\phi_i$ produces the charges of grains 
$e\delta Z_i=e(Z_i-z_i)$, the potentials of grains $\phi_i$ and  
the shift of the overall chemical potential $\delta \mu$ relative to 
$E_F$.  The condition of electroneutrality $\sum \delta Z_i=0$ yields 
\begin{eqnarray}\label{4} \phi_i&=&\frac{e\delta Z_i}{\kappa C_i},~~~ 
\delta \mu =\frac{2\alpha}{3n}\frac{\langle C_iS_i/V_i\rangle 
}{\langle C_i\rangle },\\ \delta Z_i&=& 
\frac{2\alpha\kappa}{3ne^2}\frac{C_i}{\langle C_i\rangle }(\langle 
C_iS_i/V_i\rangle - \langle C_i\rangle S_i/V_i),\label{40} 
\end{eqnarray}
where the anglular brackets denotes the mean
value. In the simple case of spherical grains
$C_i=R_i$.

According to (\ref{4},\ref{40}), the typical charge of grain is
determined by the ratio of the Fermi energy  to the
 energy of electron interaction in the metal, multiplied by $\kappa$,
and if
$R_i- \langle R\rangle\sim \langle R\rangle$, the charge $\delta Z_i$ 
may be as less, so more than unite.

 If $\alpha\kappa\ll ne^2$, so the typical $\delta Z_i\ll 1$,  the
neglecting of the charge discreetness is not valid and
the grains remain neutral.
 The transition of an electron from
grain $i$ to grain $j$ needs energy
$e^2/2\kappa~~(1/R_i+1/R_j)$.  As a result, the system of
grains proves to be the Hubbard insulator in which the
Fermi-excitations are separated by a finite gap from the ground
state.

 Controversially, if
$\alpha\kappa\gg ne^2$, the
discreetness of charge is unessential in the first approximation.
The typical potential of grain has the order of magnitude
 $E_F\lambda_F/\langle R\rangle$, and may essentially exceed the
temperature. Note, that the level spacing $\delta E$
is much less than the typical
potential, what proves the used quasiclassical approach.

Really,  (\ref{40}), gives fractional charge.
This means that  equilibrating of local electrochemical potential 
 takes place with the accuracy of the energy of 
charging of a grain by a single electron $U_C$. The 
electron redistribution proceeds as long as it is energetically 
advantageous.  The highest occupied energy level $\mu_i$ is 
separated from the overall electrochemical potential 
$E_F+\delta \mu$ by the energy, less than $U_C$, while the 
lowest empty level is above $\mu_i$ by $2U_C$. Since $|\delta Z_i|\gg 
1$, $E_F+\delta \mu-\mu_i$ are uniformly 
distributed within an energy interval $(0,U_C)$.  

Such state of
the system has zero energy gap as a whole, while the transfer of an 
electron to  a finite distance requires  finite energy. Thus, the 
metal-insulator composite may be considered as a gapless Hubbard 
insulator for $\alpha\kappa\gg ne^2$.

Let us consider the conductivity of a composite
at low temperature. In the case
$\alpha\kappa\ll ne^2$ the conductivity is determined by electrons,
excited into the upper Hubbard band (and holes in the lower band).
Hence the conductivity is purely activating with the activation
energy $min(|\pm U_c-\mu|)$, where $\mu$ is in the Hubbard
gap.

In the case of gapless Hubbard insulator the general features of
transport are similar to the hopping transport in the impurity band.

From the formal point of view, the problem is described by the
Miller-Abrahams model in which the atomic levels should be replaced
by the highest occupied levels of corresponding 
grains $\mu_i$, and the transition probability is the elastic 
tunneling probability.  The system of equations for external 
potentials of grains $\varphi_i$ and currents between grains $j_{ij}$  
has the usual form $j_{ij}=eW_{ij}(\varphi_i-\varphi_j)$, where 
\begin{eqnarray*}
-\ln W_{ij}=\zeta_{ij}=\hspace{\fill}\\2r_{ij}/a+1/2T(|\mu_{i}- 
\mu_{j}|+|\mu_{i}-\mu|+|\mu_{j}-\mu|),
\end{eqnarray*}
$r_{ij}$ is the intergrain distance, $a$ is the localization
length (the scale of electron wave function decay outside a grain).  
An electron visits the sites, where the magnitude $\zeta_{ij}$ 
satisfies the connectivity criterion $\zeta_{ij}<\zeta_c$.

Due to the continious grain spectrum  
an electron tunnels from an  exited state  of one grain 
to another grain. This process don't need phonon assistance, as 
inter-impurity transitions.  The process is determined by the density 
of grains with accessible lowest empty energy levels per unit energy 
interval $N/U_C$, instead of the integral density of states, as 
hopping transport in the impurity band. 

 At high enough temperature the tunnel factor will limit the
tunneling to the distant grains and the hops to the nearest
neighbors will prevail. The characteristic energy is determined by
some fracture of $U_c$, $\xi_cU_c$, where $\xi_c$ is the percolation
threshold.
If assume, that the grains are situated in the square or
cubic lattice,  $\xi^{(2)}_c=0.59$  and
$\xi^{(3)}_c=0.31$, correspondingly.

At low temperatures electrons prefer to jump to the distant
grains, optimizing the activation factor, resulting in the
variable range hopping mechanism of transport. The hops goes to the
grains for which the logarithms of tunneling probability and
activation have the same order.
In analogy with \cite{1}
\begin{equation}
\sigma\sim \exp-\left( \frac{T_1}{T}\right)^{1/4},~~~
 T_1\sim U_c N^{-1}a^{-3}.  \end{equation}
The Formula (\theequation) is valid within the range
$(E_F/Z)^{4/3}T_1^{-1/3}\ll T\ll U_c(Na^3)^{1/3}$. Below the lower
limit, the hopping activation energy becomes less than the distance
between single-electron levels  $E_F/Z$ and our assumption about
continuous spectrum of states fails. In this limit electrons prefer
to jump to the nearest excited levels of the nearest acceptable
grains. The typical hopping energy $\Delta$ is determined by the
condition that in the volume with radius of the hopping length $r$
there is at least one state, belonging to the infinite cluster. The
number of accessible states per unite volume in the energy range
$\Delta$ is specified by the number of states in one grain
$Z\Delta/E_F\ll 1$, multiplied by the density of grains, accessible
to the electron
$N\Delta/U_c$.
As a result, we have
\begin{equation} \label{2/5}
\sigma\sim \exp-\left( \frac{T_2}{T}\right)^{2/5},~~~
T_2\sim \left( \frac{E_F U_c}{Z Na^3}\right)^{1/2}.
\end{equation}
The equation (\ref{2/5}) obeys the law "2/5" close to the law "1/2",
observed in a number of experiments (see, for example, reviews
\cite{Abel,3,4}).
Let us emphasize, that unlike the theory of Coulomb gap,
caused by a long-range interaction of charges on different grains,
our approach takes into account more strong intra-grain charge
repulsion.   The long-range interaction has the same
order of magnitude as considered only near the percolation threshold
$NR^3\sim 1$. For Coulomb gap manifestation the temperature should be 
so low $T\lsim U_c (NR^3)^{4/9}(Na^3)^{1/3}$,
that the hopping energy to become comparable with    interaction 
of  electrons on the neighbor grains. Near the percolation 
threshold this condition transforms to $T\lsim U_c (Na^3)^{1/3}$. In 
this case the Coulomb gap is essential for transport and   leads to 
the law "1/2" \cite{1,Pollak}.  

In our consideration we neglected the collectivization of states of
different grains assuming the charging energy
$U_c$ greater than the tunnel amplitude.

We limit ourselves by the effects of the grain shape to the electron
redistribution. The other reasons of redistribution are the
fluctuations of contents (which were studied earlier in the case of
semiconductor quantum dots \cite{5}), surface states etc.

In conclusion, we evidenced, that in the system of small metal
grains they may be charged even in equilibrium at zero temperature.
The system may be Hubbard insulator with or without the energy
gap depending on whether the surface tension of electron gas is 
lower or higher than the Coulomb interaction.  The conductivity of 
the gapless insulator at low temperature is determined by the 
variable range hopping mechanism with the activation energy, caused 
by the charging energy of grains.

Authors are gratitude to B. Shklovskii for  discussions. This work was
supported by Russian Foundation for Basic Researches (Grant
97-02-18397).

\end{document}